\documentclass[a4paper]{article}

\usepackage{INTERSPEECH2020}
\usepackage[hyphens]{url}
\usepackage{
    footmisc,
    hyperref,
    colortbl,
    multirow,
}

\hypersetup{
    colorlinks=true,
    linkcolor=blue,
    citecolor=green,
    urlcolor=blue,
}

\newcommand{\x}{\vec{x}}
\newcommand{\y}{\vec{y}}
\newcommand{\z}{\vec{z}}
\newcommand{\e}{\vec{e}}

\newcommand{\D}{\mathrm{D}}
\newcommand{\G}{\mathrm{G}}
\newcommand{\sg}{\mathrm{sg}}
\newcommand{\gp}{\mathrm{gp}}

\newcommand{\E}{\mathbb{E}}

\definecolor{lightgray}{gray}{0.95}

\title{The Academia Sinica Systems of Voice Conversion for VCC2020}
\name{Yu-Huai Peng$^1$, Cheng-Hung Hu$^2$, Alexander Kang$^2$, Hung-Shin Lee$^1$, Pin-Yuan Chen$^1$, Yu Tsao$^2$, Hsin-Min Wang$^1$}

\address{
$^1$Institute of Information Science, Academia Sinica, Taiwan\\
$^2$Research Center for Information Technology Innovation, Academia Sinica, Taiwan}

\email{roland0601@iis.sinica.edu.tw}

\begin{document}

\maketitle
\begin{abstract}
This paper describes the Academia Sinica systems for the two tasks of Voice Conversion Challenge 2020, namely voice conversion within the same language (Task 1) and cross-lingual voice conversion (Task 2).
For both tasks, we followed the cascaded ASR+TTS structure, using phonetic tokens as the TTS input instead of the text or characters.
For Task 1, we used the international phonetic alphabet (IPA) as the input of the TTS model.
For Task 2, we used unsupervised phonetic symbols extracted by the vector-quantized variational autoencoder (VQVAE).
In the evaluation, the listening test showed that our systems performed well in the VCC2020 challenge.
\end{abstract}

\section{Introduction}
Voice conversion (VC) is a means of converting one voice to another. This is a technique that modifies speech waveform by converting non-linguistic information while retaining linguistic information.
While there are a wide variety of types and applications of VC, the most typical one is speaker voice conversion that converting speaker identity information while retaining linguistic information. In order to improve the VC technology, the voice conversion challenge (VCC) has been launched since 2016, and the VCC2020 challenge is the third one in the series. 

There are two tasks in VCC2020. The first task (Task 1) is VC within the same language, i.e., mono-lingual VC. The speech utterances of 4 source and 4 target speakers (consisting of both female and male speakers) from fixed corpora are used as training data. Each speaker utters a sentence set consisting of 70 sentences in English. Only 20 sentences are parallel and the other 50 sentences are nonparallel between the source and target speakers. The second task (Task 2) is cross-lingual VC. The training data includes the speech utterances of 6 target speakers (consisting of both female and male speakers) from fixed corpora and the speech utterances of the source speakers in the first task. Each target speaker utters another sentence set consisting of around 70 sentences in a different language; 2 target speakers utter in Finnish, 2 target speakers utter in German, and 2 target speakers utter in Mandarin.
Other voices of the same source speakers in English are provided later as test data consisting of around 25 sentences for each speaker. Each participant need to generate converted voices from them using the developed 16 conversion systems for the first task or 24 conversion systems for the second task.

In this paper, we describe our systems for both tasks in VCC2020. For more detailed information about VCC2020, please refer to the official website\footnote{\url{http://www.vc-challenge.org/}} and \cite{vcc2020}.   

\section{System Descriptions}
We implemented two VC systems for VCC2020, one for Task 1 and the other for Task 2.

\subsection{Task 1: voice conversion within the same language}
\begin{figure}[t]
  \includegraphics[width=0.35\textwidth]{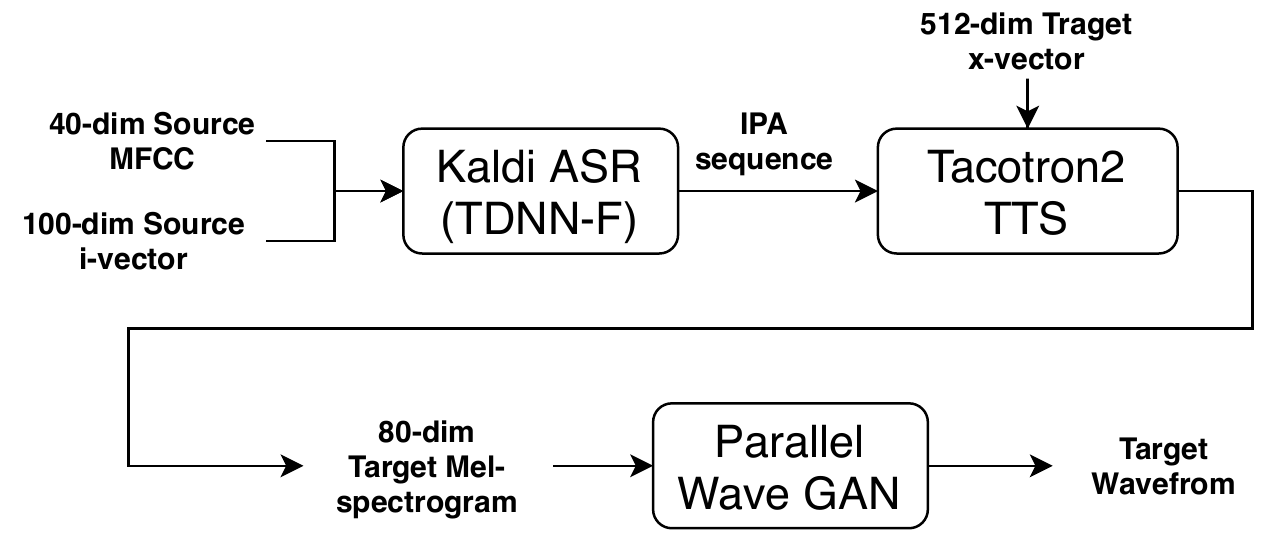}
  \centering
  \caption{
    The flow chart of our system for Task 1.
  }
  \label{fig:task1}
\end{figure}

For the first task, we built the VC system with the Kaldi ASR \cite{Povey2011} and ESPNet-TTS (Tacotron2 TTS) \cite{Hayashi2020} toolkits, as shown in Figure \ref{fig:task1}. The two models were trained independently. Finally, we used the Parallel WaveGAN \cite{Yamamoto2020} vocoder to generate waveforms to enhance naturalness and similarity. We will describe each model in the following subsections.

\subsubsection{Kaldi ASR}

To train the ASR model, we followed the Kaldi recipe\footnote{\url{https://github.com/kaldi-asr/kaldi/tree/master/egs/librispeech}} of the LibriSpeech corpus \cite{Panayotov2015}.
The training process of our ASR system was divided into two parts, data processing, and acoustic modeling.

For data processing of the LibriSpeech corpus, we first extracted two MFCC features with different resolutions.
The 13-dimensional MFCCs were used to train the GMM-based acoustic models. The 40-dimensional MFCCs were used to train the i-vector extractor and NN-based acoustic models.
Then, we used the CMU pronunciation dictionary \cite{CMUDict} to convert the English word transcriptions into the CMU pronunciation format, and mapped the CMU pronunciation symbols into the corresponding IPA symbols \cite{IPA1999}.

For acoustic modeling, the training set of the LibriSpeech corpus was first used to train the GMM-based acoustic models and the i-vector extractor. Following the model structure and training steps of the recipe, we created the alignment and lattice based on the GMM-based acoustic models, performed the data cleanup procedure, and extracted the 400-dimensional i-vectors to train the NN-based acoustic models.
For the NN-based acoustic models, we selected the ``chain'' model structure (i.e., TDNN-F) \cite{Vesely2013, Povey2016, Povey2018}. The 40-dimensional MFCCs and the 400-dimensional i-vectors were concatenated as the input of TDNN-F.

\subsubsection{Tacotron2 TTS}

To train the Tacotron2 TTS model \cite{Shen2018}, we followed the ESPNet recipe of the LibriTTS corpus\footnote{\url{https://github.com/espnet/espnet/tree/master/egs/libritts/tts1}} \cite{Zen2019}.
First, we extracted the 80-dimensional Mel-spectrogram and 512-dimensional x-vector from each utterance in the LibriTTS corpus. The speaker model used to extract the x-vectors was pre-trained with the Kaldi toolkit. As with the training of the Kaldi ASR system, the English word transcriptions were converted into the IPA symbols.

Following the model structure and training steps of the recipe, we obtained the multi-speaker Tacotron2 TTS model, which converts an IPA symbol sequence to the 80-dimensional Mel-spectrogram under the condition of a 512-dimensional x-vector. Lastly, we finetuned the Tacotron2 TTS model with the training data and average x-vector of the target speaker to obtain the speaker-dependent Tacotron2 TTS model. Note that utterance-dependent x-vectors were used to train the multi-speaker Tacotron2 TTS model, while the average  x-vector of the target speaker was used to finetune the speaker-dependent Tacotron2 TTS model. In our preliminary experiments, this combination produced the best performance.

\subsubsection{ParallelWaveGAN vocoder}

For waveform synthesis, we used ParallelWaveGAN \cite{Yamamoto2020} as the vocoder. To train the ParallelWaveGAN, similar to the ASR+TTS/ParallelWaveGAN baseline system (T22) \cite{Huang2020-2} (which uses the cascaded seq-to-seq ASR+TTS (Transformer) model for VC and ParallelWaveGAN as the vocoder), we followed the open-source ParallelWaveGAN recipe\footnote{\url{https://github.com/kan-bayashi/ParallelWaveGAN}}. We combined the VCTK corpus \cite{Veaux2012} and the training set of VCC2020, and extracted the 80-dimensional Mel-spectrogram as the input.

\subsubsection{Conversion}

In the conversion phase, the 40-dimensional MFCCs and 400-dimensional i-vectors of each input utterance were first extracted for the Kaldi ASR system to output the IPA symbol sequence. Then, the Tacotron2 TTS model of the target speaker was used to convert the IPA sequence to the 80-dimensional Mel-spectrogram of the target speaker. Finally, the ParallelWaveGAN vocoder was used to convert the 80-dimensional Mel-spectrogram of the target speaker to the waveform.

\subsection{Task 2: cross-lingual voice conversion}

\begin{figure}[t]
  \includegraphics[width=0.48\textwidth]{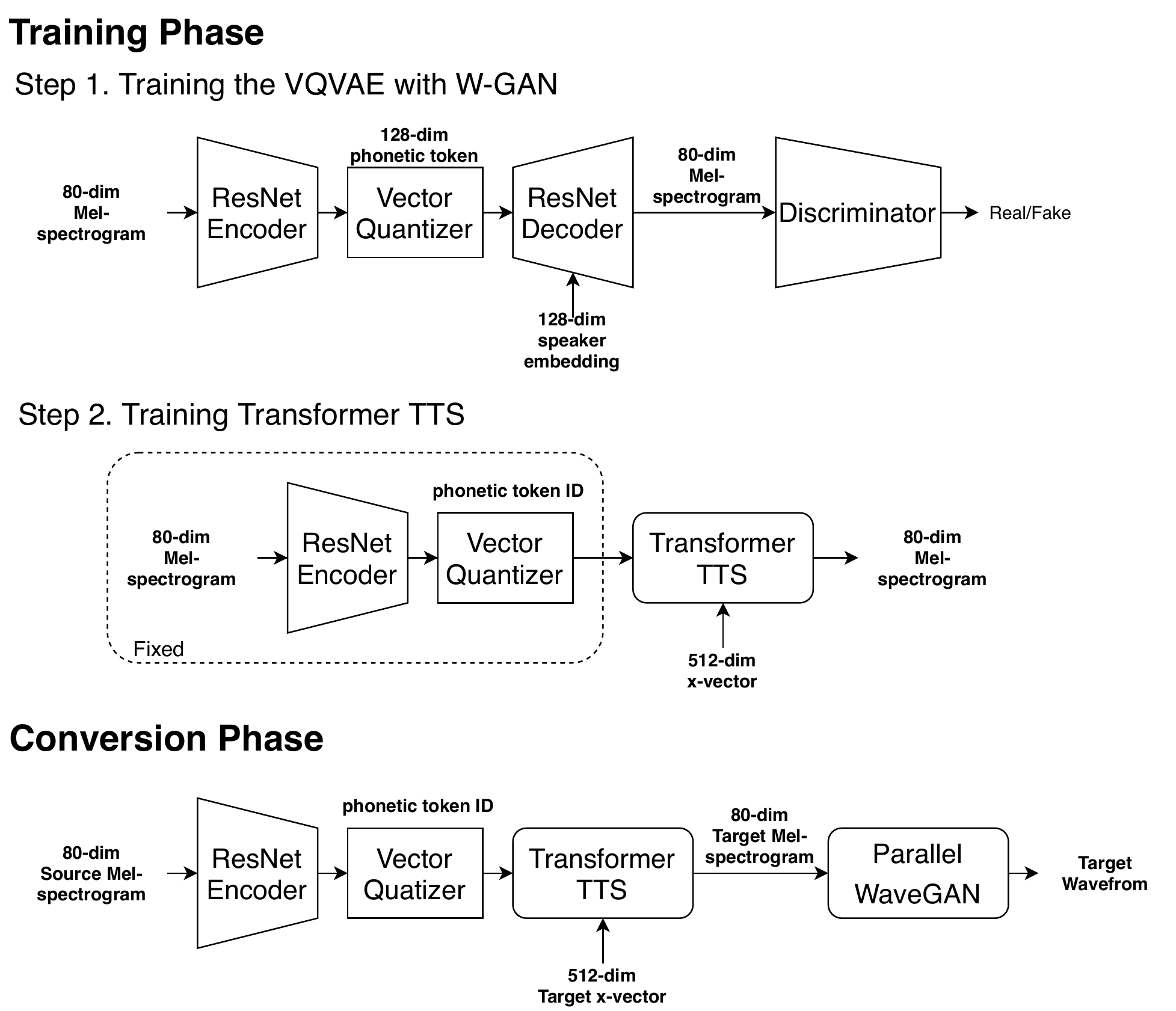}
  \centering
  \caption{
    The flow chart of our system for Task 2.
  }
  \label{fig:task2}
\end{figure}

For the second task, we built the VC system with an unsupervised phonetic symbol extractor and a Transformer TTS model \cite{Li2019}, as shown in Figure \ref{fig:task2}. Because the ASR model trained on the English corpus could not deal with non-English input speech well, we applied a variational autoencoder (VAE) based method in our system to extract the phoneme-like (or character-like) speech representations. Many studies \cite{Hsu2016, Aaron2017, Hsu2017, Qian2019, Huang2020} have shown that VAE-based methods have an ability to decompose spectral features into the speaker codes and the phonetic codes. Therefore, we applied the VQVAE \cite{Aaron2017} structure in our system to extract 
the character-like phonetic symbol sequence as the input of the Transformer TTS model. Note that we replaced the VQVAE decoder with the Transformer TTS model, because the former can only generate output with the same length as the input, while the latter can model the duration. Consequently, our system can be regarded as a seq-to-seq system.

\subsubsection{VQVAE-based phonetic symbol extractor}

Through the vector-quantization mechanism, VQVAE can quantize the latent vector representation obtained from the encoder into a discrete phonetic representation. The discrete phonetic representation is the index of the codeword closest to the latent vector in the codebook of the vector quantizer. In this way, the 80-dimensional Mel-spectrogram of an utterance is converted to a phoneme-like phonetic symbol sequence.

As shown in Figure \ref{fig:task2} (cf. Step 1), to train VQVAE, we used the W-GAN mechanism with a gradient penalty to make VQVAE perform better. The loss function $L_G$ used to update the generator of VQVAE is as follows, 
\begin{equation}
\label{eq:generator}
\begin{aligned}
    L_G = & \quad p(\x|\z_q(\x),\y) \\ 
          & + || \sg[\z_e(\x)] - \e ||^2_2 + || \z_e(\x) - \sg[\e] ||^2_2 \\
          & - \E(\D(\G(\z_q(\x),\y))),
\end{aligned}
\end{equation}
where $\x$ denotes the input feature, $\y$ denotes the speaker code, $\z_e$ denotes the latent vector representation, $\e$ denotes the closest token of the latent vector representation, $\z_q(\x) = \z_e(\x) + \sg[\e - \z_e(\x)]$ denotes the quantized discrete phonetic representation, $\sg$ denotes the stopping gradient, $\G$ denotes the decoder, and $\D$ denotes the discriminator.
The loss function $L_D$ used to update the discriminator is as follows, 
\begin{equation}
\label{eq:discriminator}
\begin{aligned}
    L_D = & - \E(\D(\x)) + \E(\D(\G(\z_q(\x),\y))) \\ 
          & + \gp(\x,\G(\z_q(\x),\y)),
\end{aligned}
\end{equation}
where $\gp$ is the gradient penalty.

We used the ResNet architecture to form the encoder, the decoder, and the discriminator. In the ResNet architecture, as shown in Tables \ref{tab:Encoder} to \ref{tab:Discriminator} , there are 4 res-layers, each containing 3 residual blocks. In the residual block of the decoder, we first concatenated the input quantized latent vector representation of size $B \times 128 \times T$ with the speaker code of size $B \times 128 \times T$ along the time axis as the new input feature, and additionally applied skip-connections, which early output skip features of size $B \times 80 \times T$.

After training the VQVAE, we only retained the encoder and vector quantizer to extract the phonetic symbol sequence from the input speech in the subsequent steps. 

\begin{table}
\caption{The architecture and specifications of Encoder, where \texttt{res1} to \texttt{res4} denote 4 ResNet-based layers, and $B$ and $T$ represent the batch size and temporal length, respectively.}
\label{tab:Encoder}
\vspace{-5pt}
\footnotesize
\centering
\begin{tabular}{cccc}
\toprule
Layer & Feature Size & Activation & Normalization \\ 
\midrule\midrule
\texttt{input}  & $B \times 80  \times T$ & - \\
\specialrule{0em}{1pt}{1pt}
\texttt{res1}   & $B \times 256 \times T$ & leaky ReLU & layer norm. \\
\specialrule{0em}{1pt}{1pt}
\texttt{res2}   & $B \times 128 \times T$ & leaky ReLU & layer norm. \\
\specialrule{0em}{1pt}{1pt}
\texttt{res3}   & $B \times 128 \times T$ & leaky ReLU & layer norm. \\
\specialrule{0em}{1pt}{1pt}
\texttt{res4}   & $B \times 128 \times T$ & leaky ReLU & layer norm. \\
\specialrule{0em}{1pt}{1pt}
\texttt{conv1d} & $B \times 128 \times T$ & None       & None \\
\bottomrule
\end{tabular}
\vspace{-0pt}
\end{table}

\begin{table}
\caption{The architecture and specifications of Decoder, where \texttt{res1} to \texttt{res4} denote 4 ResNet-based layers, \texttt{skip-sum} denotes the summation of all skip features, and $B$ and $T$ represent the batch size and temporal length, respectively.}
\label{tab:Decoder}
\vspace{-5pt}
\footnotesize
\centering
\begin{tabular}{cccccc}
\toprule
Layer & Feature Size & Activation & Normalization \\ 
\midrule\midrule
\texttt{input}  & $B \times 128  \times T$ & - \\
\specialrule{0em}{1pt}{1pt}
\texttt{res1}   & $B \times 128 \times T$ & GLU  & layer norm. \\
\specialrule{0em}{1pt}{1pt}
\texttt{res2}   & $B \times 128 \times T$ & GLU  & layer norm. \\
\specialrule{0em}{1pt}{1pt}
\texttt{res3}   & $B \times 256 \times T$ & GLU  & layer norm. \\
\specialrule{0em}{1pt}{1pt}
\texttt{res4}   & $B \times 80  \times T$ & GLU  & layer norm. \\
\specialrule{0em}{1pt}{1pt}
\texttt{skip-sum}
                & $B \times 80  \times T$ & None & None \\
\specialrule{0em}{1pt}{1pt}
\texttt{conv1d} & $B \times 80  \times T$ & GLU  & layer norm. \\
\specialrule{0em}{1pt}{1pt}
\texttt{conv1d} & $B \times 80  \times T$ & None & None \\
\bottomrule
\end{tabular}
\vspace{-0pt}
\end{table}

\begin{table}
\caption{The architecture and specifications of Discriminator, where \texttt{res1} to \texttt{res4} denote 4 ResNet-based layers, and $B$ and $T$ represent the batch size and temporal length, respectively.}
\label{tab:Discriminator}
\vspace{-5pt}
\footnotesize
\centering
\begin{tabular}{cccccc}
\toprule
Layer & Feature Size & Activation & Normalization \\ 
\midrule\midrule
\texttt{input}  & $B \times 80  \times T$ & - \\
\specialrule{0em}{1pt}{1pt}
\texttt{res1}   & $B \times 256 \times T$ & leaky ReLU & layer norm. \\
\specialrule{0em}{1pt}{1pt}
\texttt{res2}   & $B \times 128 \times T$ & leaky ReLU & layer norm. \\
\specialrule{0em}{1pt}{1pt}
\texttt{res3}   & $B \times 64  \times T$ & leaky ReLU & layer norm. \\
\specialrule{0em}{1pt}{1pt}
\texttt{res4}   & $B \times 32  \times T$ & leaky ReLU & layer norm. \\
\specialrule{0em}{1pt}{1pt}
\texttt{conv1d} & $B \times 1   \times T$ & None       & None \\
\bottomrule
\end{tabular}
\vspace{-10pt}
\end{table}

\subsubsection{Transformer TTS}

To train the Transformer TTS model, we followed the ESPNet recipe of the LibriTTS corpus, but replaced the training corpus with a combination of the VCTK corpus and the VCC2020 training set.
First, we extracted the 80-dimensional Mel-spectrogram and 512-dimensional x-vector from each utterance in the training corpus. The speaker model used to extract the x-vectors was pre-trained with the Kaldi toolkit. The phoneme-like phonetic symbol sequence of each training utterance was extracted using the VQVAE encoder and vector quantizer.

Following the model structure and training steps of the recipe, we obtained the multi-speaker Transformer TTS model, which convert the phoneme-like phonetic symbol sequence to the 80-dimensional Mel-spectrogram under the condition of a 512-dimensional x-vector. Lastly, we finetuned the Transformer TTS model with the training data and average x-vector of the target speaker to obtain the speaker-dependent Transformer TTS model.

\subsubsection{Conversion}

In the conversion phase, the 80-dimensional Mel-spectrogram of each source utterance was first passed to the VQVAE encoder and vector quantizer  to generate the phoneme-like phonetic symbol sequence. Then, the Transformer TTS model of the target speaker was used to convert the phoneme-like phonetic symbol sequence to the 80-dimensional Mel-spectrogram of the target speaker. Finally, the ParallelWaveGAN vocoder was used to convert the 80-dimensional Mel-spectrogram of the target speaker to the waveform.

\section{Experiment Results}
\label{sec:experiments}

As stated in the final report of VCC2020 \cite{vcc2020}, all submitted systems were grouped according to their performance. The systems in each group did not differ significantly in performance. 
According to the evaluation results of VCC2020 \cite{vcc2020}, our system for Task 1 ranked in the fifth group in terms of naturalness (31 systems were ranked and divided into 18 groups). In terms of similarity to the target speaker, our system for Task 1 ranked in the first group (31 systems were divided into 9 groups). For Task 2, our system ranked in the fifth group in terms of naturalness (28 systems were divided into 15 groups) and ranked in the sixth group in terms of similarity to the target speaker (28 systems were divided into 13 groups).

For Task 1, we presented a new ASR+TTS system. In recent studies, the ASR+TTS systems achieved good performance. Therefore, we tried to make some improvements on the basis of the baseline ASR+TTS system (T22) \cite{Huang2020-2}. We built our ASR system based on the IPA symbols. Our goal was not only to use this ASR+TTS system to accomplish Task 1, but also to apply the same ASR+TTS system to Task 2. However, we found that the model could not have consistent performance in some VC pairs in the cross-lingual VC task. One possible reason is that we did not have enough training data to train our ASR+TTS system for the cross-lingual VC task.
It turned out that our ASR+TTS system performed as well as the baseline ASR+TTS system (T22) in the mono-lingual VC task. Note that the baseline ASR+TTS system is a cascade of seq-to-seq ASR and Transformer TTS models implemented using the end-to-end speech processing toolkit ``ESPNet'' \cite{Hayashi2020}  \footnote{\url{https://github.com/espnet/espnet/tree/master/egs/vcc20}}.   
According to the evaluation results of VCC2020, in Task 1, our system roughly ranked in the top 30\% in terms of naturalness and similarity. 

For Task 2, we modified the traditional VQVAE VC system and replaced the decoder with a Transformer TTS model. In our preliminary experiments, we found that replacing the decoder with the Transformer TTS model could improve the naturalness, but the similarity was almost the same. This result is in line with the VCC2020 evaluation results. Comparing our system with the two VQVAE-based systems (T19 and T20) in \cite{vcc2020}, we can see that in naturalness, our system is comparable to T20, but better than T19; while in similarity, our system is worse than T19, but better than T20. In addition, comparing our system with the VAE-based baseline system (T16), i.e., the CycleVAE VC system with ParallelWaveGAN as the vocoder \cite{Tobing2019, Tobing2020}, we can see that our system is better than T16 in naturalness, but worse than T16 in similarity. In the naturalness test, our system was ranked in the fifth group with the MOS score of 3.00, while the baseline T16 system was ranked in the ninth group with the MOS score of 2.56. In the similarity test, our system was ranked in the sixth group with the score of 2.41, while the baseline T16 system was ranked in the fourth group with the score of 2.69. 
In the challenge, our MOS score in Task 2 was in the upper range, about the top 30\%. However, our system ranked in the middle in terms of similarity. In terms of ranking, our Task 2 system was not as good as our Task 1 system. There are two possible reasons. First, we did not optimize the VQVAE encoder to suit the task. Second, the vanilla VQVAE model we used has its own performance limitations. We will try to improve our system on these two issues in the future.

\section{Conclusions}

Ideally, the ASR+TTS model can perfectly retain the linguistic information and synthesize the linguistic information with new personal identity information into the target speech.
However, according to our preliminary  experiments, the ASR+TTS model only performed well in mono-lingual VC tasks, but not in cross-lingual VC tasks. Therefore, we built an alternative VQVAE+TTS model for Task 2. We expected that the encoder of the VQVAE model could replace the role of ASR. The difference between VQVAE encoder and ASR is that the output of ASR is a sequence of symbols defined by human, such as words and phones (in IPA or phonetic posteriograms), while the output of VQVAE encoder is a sequence of tokens automatically learned by the machine. The codewords in the codebook learned by the VQ part in VQVAE could be regarded as recognition units in the ASR model. According to our experiments, the VQVAE+TTS model did achieve better performance than the ASR+TTS model in the cross-lingual task. However, as discussed in Section \ref{sec:experiments}, there are still some problems to be solved.


\bibliographystyle{IEEEtran}
\bibliography{references.bib}

\begin{thebibliography}{10}
\providecommand{\url}[1]{#1}
\csname url@samestyle\endcsname
\providecommand{\newblock}{\relax}
\providecommand{\bibinfo}[2]{#2}
\providecommand{\BIBentrySTDinterwordspacing}{\spaceskip=0pt\relax}
\providecommand{\BIBentryALTinterwordstretchfactor}{4}
\providecommand{\BIBentryALTinterwordspacing}{\spaceskip=\fontdimen2\font plus
\BIBentryALTinterwordstretchfactor\fontdimen3\font minus
  \fontdimen4\font\relax}
\providecommand{\BIBforeignlanguage}[2]{{%
\expandafter\ifx\csname l@#1\endcsname\relax
\typeout{** WARNING: IEEEtran.bst: No hyphenation pattern has been}%
\typeout{** loaded for the language `#1'. Using the pattern for}%
\typeout{** the default language instead.}%
\else
\language=\csname l@#1\endcsname
\fi
#2}}
\providecommand{\BIBdecl}{\relax}
\BIBdecl

\bibitem{vcc2020}
Y.~Zhao, W.-C. Huang, X.~Tian, J.~Yamagishi, R.~K. Das, T.~Kinnunen, Z.~Ling,
  and T.~Toda, ``Voice conversion challenge 2020: Intra-lingual semi-parallel
  and cross-lingual voice conversion,'' in \emph{Proc. ISCA Joint Workshop for
  the Blizzard Challenge and Voice Conversion Challenge 2020}.\hskip 1em plus
  0.5em minus 0.4em\relax ISCA, 2020.

\bibitem{Povey2011}
D.~Povey, A.~Ghoshal, G.~Boulianne, L.~Burget, O.~Glembek, N.~Goel,
  M.~Hannemann, P.~Motl{\'{i}}{\v{c}}ek, Y.~Qian, P.~Schwarz,
  J.~Silovsk{\'{y}}, G.~Stemmer, and K.~Vesel, ``{The Kaldi speech recognition
  toolkit},'' in \emph{Proc. ASRU}, 2011.

\bibitem{Hayashi2020}
T.~Hayashi, R.~Yamamoto, K.~Inoue, T.~Yoshimura, S.~Watanabe, T.~Toda,
  K.~Takeda, Y.~Zhang, and X.~Tan, ``{ESPnet-TTS: Unified, reproducible, and
  integratable open source end-to-end text-to-speech toolkit},'' in \emph{Proc.
  ICASSP}, 2020.

\bibitem{Yamamoto2020}
R.~Yamamoto, E.~Song, and J.~M. Kim, ``{Parallel WaveGAN: A fast waveform
  generation model based on generative adversarial networks with
  multi-resolution spectrogram},'' in \emph{Proc. ICASSP}, 2020.

\bibitem{Panayotov2015}
V.~Panayotov, G.~Chen, D.~Povey, and S.~Khudanpur, ``{LibriSpeech: An ASR
  corpus based on public domain audio books},'' in \emph{Proc. ICASSP}, 2015.

\bibitem{CMUDict}
\BIBentryALTinterwordspacing
{Carnegie Mellon University}, \emph{{CMU pronouncing dictionary.}} [Online].
  Available: \url{http://www.speech.cs.cmu.edu/cgi-bin/cmudict}
\BIBentrySTDinterwordspacing

\bibitem{IPA1999}
{International Phonetic Association}, \emph{{Handbook of the International
  Phonetic Association: A Guide to the Use of the International Phonetic
  Alphabet}}, 1999.

\bibitem{Vesely2013}
K.~Vesely, A.~Ghoshal, L.~Burget, and D.~Povey, ``{Sequence-discriminative
  training of deep neural networks},'' in \emph{Proc. Interspeech}, 2013.

\bibitem{Povey2016}
D.~Povey, V.~Peddinti, D.~Galvez, P.~Ghahrmani, V.~Manohar, X.~Na, Y.~Wang, and
  S.~Khudanpur, ``{Purely sequence-trained neural networks for ASR based on
  lattice-free MMI},'' in \emph{Proc. Interspeech}, 2016.

\bibitem{Povey2018}
D.~Povey, G.~Cheng, Y.~Wang, K.~Li, H.~Xu, M.~Yarmohamadi, and S.~Khudanpur,
  ``{Semi-orthogonal low-rank matrix factorization for deep neural networks},''
  in \emph{Proc. Interspeech}, 2018.

\bibitem{Shen2018}
J.~Shen, R.~Pang, R.~J. Weiss, M.~Schuster, N.~Jaitly, Z.~Yang, Z.~Chen,
  Y.~Zhang, Y.~Wang, R.~Skerrv-Ryan, R.~A. Saurous, Y.~Agiomvrgiannakis, and
  Y.~Wu, ``{Natural TTS synthesis by conditioning WaveNet on Mel spectrogram
  predictions},'' in \emph{Proc. ICASSP}, 2018.

\bibitem{Zen2019}
H.~Zen, V.~Dang, R.~Clark, Y.~Zhang, R.~J. Weiss, Y.~Jia, Z.~Chen, and Y.~Wu,
  ``{LibriTTS: A corpus derived from LibriSpeech for text-to-speech},'' 2019.

\bibitem{Huang2020-2}
W.-C. Huang, T.~Hayashi, S.~Watanabe, and T.~Toda, ``{The Sequence-to-Sequence
  Baseline for the Voice Conversion Challenge 2020: Cascading ASR and TTS},''
  in \emph{Proc. ISCA Joint Workshop for the Blizzard Challenge and Voice
  Conversion Challenge 2020}.\hskip 1em plus 0.5em minus 0.4em\relax ISCA,
  2020.

\bibitem{Veaux2012}
\BIBentryALTinterwordspacing
C.~Veaux, J.~Yamagishi, and K.~MacDonald, ``{CSTR VCTK corpus: English
  multi-speaker corpus for CSTR voice cloning toolkit},'' 2012. [Online].
  Available: \url{https://doi.org/10.7488/ds/1994}
\BIBentrySTDinterwordspacing

\bibitem{Li2019}
N.~Li, S.~Liu, Y.~Liu, S.~Zhao, and M.~Liu, ``{Neural speech synthesis with
  Transformer network},'' in \emph{Proc. AAAI Conference on Artificial
  Intelligence}, 2019.

\bibitem{Hsu2016}
C.~C. Hsu, H.~T. Hwang, Y.~C. Wu, Y.~Tsao, and H.~M. Wang, ``{Voice conversion
  from non-parallel corpora using variational auto-encoder},'' in \emph{Proc.
  APSIPA ASC}, 2016.

\bibitem{Aaron2017}
A.~{Van Den Oord}, O.~Vinyals, and K.~Kavukcuoglu, ``{Neural discrete
  representation learning},'' in \emph{Proc. NIPS}, 2017.

\bibitem{Hsu2017}
C.~C. Hsu, H.~T. Hwang, Y.~C. Wu, Y.~Tsao, and H.~M. Wang, ``Voice conversion
  from unaligned corpora using variational autoencoding wasserstein generative
  adversarial networks,'' in \emph{Proc. Interspeech}, 2017.

\bibitem{Qian2019}
K.~Qian, Y.~Zhang, S.~Chang, X.~Yang, and M.~Hasegawa-Johnson, ``{AUTOVC:
  Zero-shot voice style transfer with only autoencoder Loss},'' in \emph{Proc.
  ICML}, 2019.

\bibitem{Huang2020}
W.~C. Huang, H.~Luo, H.~T. Hwang, C.~C. Lo, Y.~H. Peng, Y.~Tsao, and H.~M.
  Wang, ``{Unsupervised representation disentanglement using cross domain
  features and adversarial learning in variational autoencoder based voice
  conversion},'' \emph{IEEE Transactions on Emerging Topics in Computational
  Intelligence}, vol.~4, no.~4, pp. 468--479, 2020.

\bibitem{Tobing2019}
P.~L. Tobing, Y.-C. Wu, T.~Hayashi, K.~Kobayashi, and T.~Toda, ``{Non-Parallel
  Voice Conversion with Cyclic Variational Autoencoder},'' in \emph{Proc.
  Interspeech}, 2019.

\bibitem{Tobing2020}
P.~L. Tobing, Y.-C. Wu, and T.~Toda, ``The baseline system of voice conversion
  challenge 2020 with cyclic variational autoencoder and parallel wavegan,'' in
  \emph{Proc. ISCA Joint Workshop for the Blizzard Challenge and Voice
  Conversion Challenge 2020}.\hskip 1em plus 0.5em minus 0.4em\relax ISCA,
  2020.

\end{thebibliography}

\end{document}